\documentclass[aps,prb,showpacs,twocolumn]{revtex4}
\usepackage{graphicx}
\usepackage{amsmath}
\usepackage{color}
\usepackage{amsbsy}

\def\bA{{\bf A}}
\def\bE{{\bf E}}
\def\cH{\hat{\cal H}}
\def\cL{{\cal L}}
\def\bp{{\bf p}}
\def\br{{\bf r}}
\def\bv{{\bf v}}

\def\bsigma{{\boldsymbol\sigma}}

\input{epsf}
\begin{document}

\title{Photocurrent in a visible-light graphene photodiode}

\author{S.~Mai, S.V.~Syzranov, and K.B.~Efetov}
\affiliation{Theoretische Physik III, Ruhr-Universit\"at Bochum,
44780 Bochum, Germany}
\date{\today}

\begin{abstract}
We calculate the photocurrent in a clean graphene sample normally
irradiated by a monochromatic electromagnetic field and subject to a
step-like electrostatic potential. We consider the photon energies
$\hbar\Omega$ that significantly exceed the height of the potential
barrier, as is the case in the recent experiments with
graphene-based photodetectors. The photocurrent comes from the
resonant absorption of photons by electrons and decreases with
increasing ratio $\hbar\Omega/U_0$. It is weakly affected by the
background gate voltage and depends on the light polarization as
$\propto\sin^2\gamma$, $\gamma$ being the angle between the
potential step and the polarization plane.
\end{abstract}

\pacs{72.80.Vp, 78.67.Wj, 85.60.-q, 73.23.Ad}


\maketitle


Unique transport and optical properties of graphene make it likely
to find a broad application in
optoelectronics\cite{Bonaccorso:photoreview}. Those include, in
particular, a remarkable purity of this two-dimensional
semiconductor and its gapless bandstructure, that enables one to
easily change the doping level by applying gate voltages and operate
graphene devices in a broad range of external radiation. Unlike the
case of ordinary semiconductors, the frequency of applied radiation
may be rather low, owing to the absence of forbidden band in
graphene. Apart from practical applications, graphene reveals a
bunch of new fundamental light-induced phenomena, for example the
photon-assisted interference between electron
paths\cite{Fistul:interference} or the Hall effect without magnetic
field\cite{Hall}.

Over the past year there have been demonstrated a number of
graphene-based photodetectors\cite{Xia:photoimaging,
Xia:photodetector, Park:multigate, Xu:thermojunction}. The simplest
device of this kind is an irradiated graphene sample subject to a
step-like electrostatic potential, for example a $p-n$ junction or a
unipolar ($n-n$ or $p-p$) junction. Such a detector can operate in a
wide frequency range of external radiation for there is no gap
between the conduction and the valence bands in graphene. To achieve
the maximal photocurrent one should apply radiation with the photon
energy $\hbar\Omega$ of the order of the height of the potential
barrier $U_0$ in the junction\cite{Syzranov:photocurrent}. As the
attainable doping level in graphene lies within hundreds of
millivolts, this corresponds to the radiation in the terahertz or
the far-infrared frequency range.

Transport in illuminated graphene junctions with the ratio
$\hbar\Omega/U_0$ of order unity has been analyzed in detail in
Ref.~\onlinecite{Syzranov:photocurrent}. However, the more
experimentally accessible radiation wavelengths, yet those of
greater practical importance, belong to the near-infrared and the
visible range, corresponding to the photon energies $\hbar\Omega$
significantly exceeding the characteristic electrostatic potentials
that can be created in graphene devices by means of gate electrodes.
The analysis of the photocurrent in the latter regime (i.e., at
$\hbar\Omega/U_0\gg1$) is the subject of the present Brief Report.

Generally speaking, photocurrent in an irradiated graphene sample
may arise due to a number of reasons. Even in absence of external
potential it may be caused by the photon drag effect or by the
light-induced currents on the edges of the sample\cite{Ganichev:ext,
Entin:drag}.
These two mechanisms of generating the photocurrent can be separated
from the others by checking the dependency of the current on the
angle of incidence and by moving the light spot to the edge of the
sample, respectively\cite{Ganichev:ext}. If a sample is heated
nonuniformly by the radiation, the photocurrent may also arise due to the thermophotoelectric
effect\cite{Xu:thermojunction}. Clearly, the photocurrent is not
allowed by the symmetry in a normally irradiated uniform sample with
the borders equally (un)affected by the light.

In the recent experiments with graphene photodetectors (Refs.~\onlinecite{Xia:photoimaging, Xia:photodetector}) the voltages
on the gate electrodes determine the value of the photocurrent, the
latter being a direct measure of the slope of the potential profile.
This suggests that the light absorption leads to the creation of
electron-hole pairs, separated further by the electric field in the
junction, which results in the generation of the photocurrent.

\begin{figure}[t]
\includegraphics[width=1.8in,angle=0]{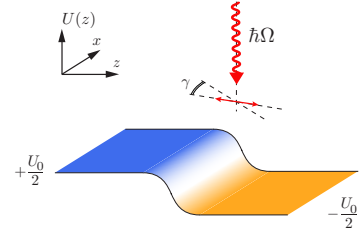}
\caption{\label{setup} (Colour online) Graphene junction irradiated
by an electromagnetic wave.} \label{Schematic}
\end{figure}

In the present Brief Report we calculate the photocurrent that emerges in
irradiated graphene sample in presence of a nonuniform potential
due to the resonant absorption of photons by electrons. We consider
a wide graphene strip (Fig.~\ref{setup}) subject to a smooth
potential $U(z)$ which varies monotonously from $U_0/2$ at
$z=-\infty$ (left lead) to $-U_0/2$ at $z=+\infty$ (right lead).

{\it Result.} We find the photocurrent as
\begin{eqnarray}
   I=\kappa \frac{e^3W}{\hbar
   cF}\left(\frac{U_0}{\hbar\Omega}\right)^{\frac{3}{2}}S\sin^2\gamma,
   \label{result1}
\end{eqnarray}
where $\gamma$ is the angle between polarization plane of the light
and the potential barrier (axis $x$ in Fig.~\ref{setup}), $S$ is the
radiation intensity, $W$ is the width of the strip, $F$ is the
characteristic slope of the potential, and $\kappa$ is a constant of
order unity.

Equation~(\ref{result1}) indicates that the photocurrent vanishes if the
polarization plane is parallel to the barrier ($\gamma=0$). In fact,
it does not vanish completely, but acquires an extra power of the
small parameter $U_0/\hbar\Omega\ll1$,
\begin{eqnarray}
   I_\|=\kappa_\| \frac{e^3W}{\hbar
   cF}\left(\frac{U_0}{\hbar\Omega}\right)^{\frac{5}{2}}S,
   \label{result2}
\end{eqnarray}
$\kappa_\|$ is another constant of order unity.

Coefficients $\kappa$ and $\kappa_\|$ can be evaluated exactly for
any particular form of the potential barrier. For instance, if the
slope $F$ of the potential is constant on the interval from
$z=-U_0/(2F)$ to $z=U_0/(2F)$ and zero otherwise, then
$\kappa=8/3$ and $\kappa_\|=32/15$.

{\it Model.} The Hamiltonian of electrons in irradiated graphene in
each valley reads
\begin{eqnarray}\label{Hamiltonian}
    \cH=v\hat\bsigma\left[\bp-{e}{c}^{-1}\bA(t)\right]+U(z),
\end{eqnarray}
where the vector potential $\bA(t)$ accounts for the external
electromagnetic field (EF). For a linearly polarized wave one can choose
\begin{eqnarray}
    \bA(t)={c}{\Omega}^{-1}\bE\cos(\Omega t).
\end{eqnarray}
The ``pseudospin'' $\bsigma$ in Eq.~(\ref{Hamiltonian}) is the
spin-1/2 operator defined on the space of the two sublattices in
graphene. The characteristic size of the step-like barrier lies
typically within dozens of nanometers (see, e.g.,
Ref.~\onlinecite{Xia:photoimaging}), not exceeding the mean free
path, so the transport in absence of radiation may be considered
ballistically.

{\it Scattering off photons.} The dynamics of electrons affected by
light has been considered microscopically in
Ref.~\onlinecite{Syzranov:photocurrent}. We provide here a simple
semiqualitative picture describing the main features of this
dynamics.

The light strongly affects electron motion only close to the
``resonant points'', where the splitting $2vp$ between the
conduction and valence bands matches the photon energy
$\hbar\Omega$. Far from the resonant points the radiation weakly
affects electron dynamics and can be neglected. The motion between
the resonant points can be considered semiclassically. Electron
velocity there has a constant value $v$, as follows from
Eq.~(\ref{Hamiltonian}) at $\bA=0$.

The rate of the radiation-induced transitions between the conduction
and the valence bands,
\begin{eqnarray}\label{transrate}
    \Gamma={2\pi}{\hbar}^{-1}\Delta^2\sin^2\beta\:\delta(2vp-\hbar\Omega),
\end{eqnarray}
may be viewed at small radiation intensities as a mere
Fermi-golden-rule result. Here the ``dynamical
gap''\cite{Syzranov:photocurrent}
\begin{eqnarray}
\Delta={v|e|E}/({2\Omega})
\end{eqnarray}
characterizes the strength of the radiation and $\beta$ is the
angle between the electron momentum and the light polarization
plane. The delta function guarantees that the transitions occur only
at the resonant points. During the light-induced scattering,
electron momentum $\bp$ does not change, but the pseudospin flips
and the energy increases (decreases) by $\hbar\Omega$ when absorbing
(emitting) a photon.

Let $\alpha$ be the angle between the classical electron momentum
$\bp$ and the potential gradient $dU/d\br$. Integrating
Eq.~(\ref{transrate}) over time and using that $\dot{\bp}=-dU/d\br$
we find the probability of electron scattering at the resonant point
\begin{eqnarray}\label{transprob}
    \cL(\beta,\alpha)={\pi\Delta^2\sin^2\beta}/({vF\cos\alpha}).
\end{eqnarray}
 If an
electron runs against a resonant point when moving along a certain
classical trajectory, it scatters with probability $\cL$ or
continues its motion undisturbed by the radiation along the same
trajectory with probability $1-\cL$.

The above results, Eqs.~(\ref{transrate}) and (\ref{transprob}), are
obtained perturbatively in the limit of small $\Delta$ and thus
require sufficiently small radiation powers. One can also derive
Eqs.~(\ref{transrate}) and (\ref{transprob}) explicitly by solving
the Schr{\"o}dinger equation for electrons in presence of EF or
using the kinetic equation\cite{Syzranov:photocurrent}. The
condition of smallness of the radiation power reads $\cL\ll1$ and
corresponds to the most experimentally relevant range of the system
parameters.

{\it Formula for the current.} The previous generic picture of the
light-affected electron dynamics in a nonuniform potential allows
one to find the electron trajectories and to calculate the
photocurrent as\cite{Syzranov:photocurrent}
\begin{equation}
    I=4W\sum_n\int\frac{d\bp}{(2\pi\hbar)^2}\bv_{\|}P_{n}(\bp)
    \left\{f[\varepsilon(\bp)]-f[\varepsilon(\bp)+n\hbar\Omega]\right\},
    \label{Landauer}
\end{equation}
where the integration is carried out over the incoming electron
momenta $\bp$ in the left lead, $\bv_{\|}$ is the respective
longitudinal velocity, $P_{n}(\bp)$ the probability for an electron
outgoing from the left lead to penetrate into the right lead
absorbing $n$ photons, $f(\varepsilon)$ is the equilibrium
distribution function in both leads, and $W$ the width of the graphene
strip. Equation~(\ref{Landauer}) is, in fact, the Landauer formula generalized
to account for the inelastic processes of photon
absorption/emission.

In principle, our scheme of calculations follows that of
Ref.~\onlinecite{Syzranov:photocurrent}: We have to count all the
classical trajectories corresponding to the inelastic electron
transmission from the left to the right lead and then, using
Eq.~(\ref{Landauer}), calculate the value of the photocurrent. {In
the case of a shallow potential barrier, $U_0\ll\hbar\Omega$,
studied in the present Brief Report, we are dealt with a greater variety of
electron paths than in the previously studied case of a relatively
high potential\cite{Syzranov:photocurrent}. Let us consider these
trajectories in detail.}

\begin{figure}[t!]
\includegraphics[height=1.4in,angle=0]{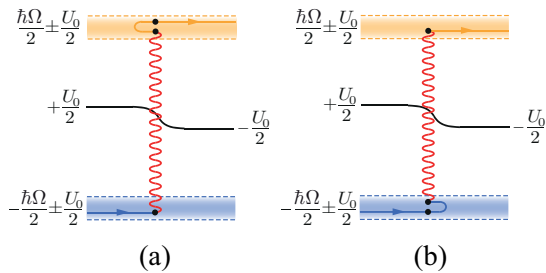}
\caption{\label{trajectories2} (Colour online) Inelastic penetration
of an electron from the left to the right lead. The reflection from
the potential barrier must occur either before (a) or after (b) the
photon absorption.}
\end{figure}

{\it Electron trajectories.} As is follows from the resonance
condition $\hbar\Omega=2vp$, the kinetic energies of an electron
before (after) and after (before) the photon absorption (emission)
equal, respectively, $vp=\hbar\Omega/2$ and $vp=-\hbar\Omega/2$.
Hence photons can be absorbed only by incident electrons in the
narrow energy band (Fig.~\ref{trajectories2})
\begin{eqnarray}
    -{\hbar\Omega}/{2}-{U_0}/{2}<\varepsilon<-{\hbar\Omega}/{2}+{U_0}/{2},
    \label{absorptioncondition}
\end{eqnarray}
far below the Fermi level. So long as the latter has the same order
of magnitude as $U_0$, its exact position is not important for the
photocurrent.

\begin{figure}[t!]
\includegraphics[width=.9\columnwidth,angle=0]{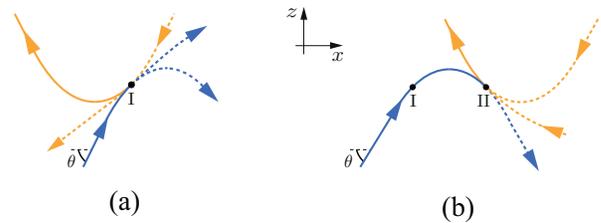}
\caption{\label{trajectories1} (Colour online) Classical
trajectories of electrons incident on the potential barrier. Photon
absorption occurs at (a) the first resonant point (b) the second resonant
point. Solid lines show the paths that contribute to the
photocurrent, dashed lines show the other possible trajectories. Blue
(black) and orange (gray) lines correspond, respectively, to the
motion in the valence and the conduction bands.}
\end{figure}

Each electron with energy in the interval
(\ref{absorptioncondition}) and with a sufficiently small transverse
momentum
\begin{eqnarray}
    vp_\bot<{\hbar\Omega}/{2},
    \label{transversesmall}
\end{eqnarray}
 inevitably meets a resonant point on its way from one lead
to the other. Assume an electron incident from the left lead
absorbs a photon at the very first resonant point I,
Fig.~\ref{trajectories1}(a). Before and after the absorption, its
longitudinal velocity is directed, respectively, to and from the right
lead. The velocity reversal may result in the return to the
left lead, as shown by the orange (gray) line in
Fig.~\ref{trajectories1}(a). The electron, however, may also penetrate
into the right lead and thus contribute to the photocurrent if it is
reflected from the potential barrier after the photon absorption
[see the orange (gray) solid line in Fig.~\ref{trajectories1}(a)]. As
the transverse momentum $p_\bot$ is conserved during the motion,
$\varepsilon+\hbar\Omega$ and $U_0/2$ are, respectively, the full
energy and the potential in the left lead, the return to this lead
does not occur if and only if
\begin{eqnarray}
    \varepsilon+\hbar\Omega-vp_\bot<{U_0}/{2}.
    \label{conditionafterreflection}
\end{eqnarray}

Thus, we have shown that the charge carriers, which absorb photons
at their first resonant points and satisfy conditions
(\ref{absorptioncondition})-(\ref{conditionafterreflection}),
contribute to the photocurrent.

What if the photon absorption at the first resonant point did not
happen? Then the electron can proceed further to the other lead,
meeting no other resonant points, or it can turn back,
Fig.~\ref{trajectories1}, provided
\begin{eqnarray}
    -\varepsilon-U_0<vp_\bot.
    \label{reflection}
\end{eqnarray}
 In the latter case the second
resonant point II will be reached, for the longitudinal momentum
decreases monotonously up to a certain turning point and then grows
again. The sign of the longitudinal velocity at the second resonant
point is opposite to that at the first one, so after the reflection
the electron moves towards the right lead, along the orange (gray)
line in Fig.~\ref{trajectories1}(b). Then, since the momentum of the
electron grows monotonously, neither resonant nor turning points can
be met further.

Therefore, each electron, whose energy and momenta satisfy
conditions (\ref{absorptioncondition}), (\ref{transversesmall}), and
(\ref{reflection}) can contribute to the photocurrent upon the
photon absorption at point II.

Clearly, elastic penetration from one lead to another is also
possible, but, according to Eq.~(\ref{Landauer}), does not
contribute to the photocurrent. We have considered then all the
scenarios of the light-assisted transmission from left to right, the
contribution with $n=+1$ in Eq.~(\ref{Landauer}). In principle,
 we are also to deal with electrons
that absorb a photon on their way from right to left. It is more
convenient, however, to consider the time-reversed processes and
speculate in terms of the states outgoing from the same left lead
and emitting a photon at the resonant points. Indeed, in
Eq.~(\ref{Landauer}) these processes are accounted for by the terms
with $n=-1$, whereas the integration is carried out over the states
in the left lead only.

The energies of the charge carriers, that can emit photons, lie in
the range
\begin{eqnarray}
    {\hbar\Omega}/{2}-{U_0}/{2}<\varepsilon<{\hbar\Omega}/{2}+{U_0}/{2}.
    \label{conductioninterval}
\end{eqnarray}
The reflection from the potential barrier without emitting a photon
is not possible, for the longitudinal momentum of the electrons in
the conduction band grows monotonously as the potential decreases.
This forbids the processes analogous to what is shown in
Fig.~\ref{trajectories1}(b). Similarly, if an emission occurs at a
resonant point, electron inevitably returns to the left lead.

Thus, there is no contribution to the photocurrent from the energy
interval (\ref{conductioninterval}), and one has to take into
account only the processes shown by solid lines in
Fig.~\ref{trajectories1}. Each of this processes involves at least
one piece of a classical trajectory reflecting from the potential
barrier. In the shallow potential under consideration the reflection
is possible only during the motion nearly parallel to the barrier
(axis $x$). {Hence, the photocurrent found in the present Brief Report
should be significantly smaller than that studied before in
Ref.~\onlinecite{Syzranov:photocurrent} at large ratio
$U_0/\hbar\Omega$. In the latter case nearly all electrons rebound
from the barrier and photon absorption assists hopping from one
trajectory of reflecting type to another such trajectory,
similarl to what is shown in Fig.~\ref{trajectories1} by the solid
lines. As we have shown, at $U_0/\hbar\Omega\ll1$ the structure of
electron paths is more diverse and very few of them contribute to
the photocurrent. }

{\it Integration over the electron states.} We must take into
account the contributions to the photocurrent of electrons with
momenta $p$ in the left lead, such that
\begin{eqnarray}
    {\hbar\Omega}/({2v})<p<{\hbar\Omega}/({2v})+{U_0}/{v},
    \label{momentainterv}
\end{eqnarray}
[cf., Eq.~(\ref{absorptioncondition})] and incident at angles $\theta$
between the momenta $\bp$ and the barrier that lie in one of the two
intervals
\begin{eqnarray}
    \sqrt{2}\left[1-{\hbar\Omega}/({2vp})\right]^{{1}/{2}}<\theta
    <2\left[1-{\hbar\Omega}/({2vp})\right]^{{1}/{2}},
    \label{angleinterv1}
    \\
    \sqrt{2}\left[1-{\hbar\Omega}/({2vp})\right]^{{1}/{2}}<\theta
    <\left[{2U_0}/({vp})\right]^{{1}/{2}},
    \label{angleinterv2}
\end{eqnarray}
corresponding, respectively, to Figs.~\ref{trajectories1}(a) and
\ref{trajectories1}(b).

The photon absorption at low radiation powers occurs with small
probability
\begin{eqnarray}
    \cL={\pi\Delta^2\sin^2\gamma}/{\theta_{\mathrm{res}}},
    \label{Lactual}
\end{eqnarray}
where $\theta_{\mathrm{res}}$ is the angle between the momentum and the
barrier at the resonant point, determined by the transverse momentum
conservation law
\begin{equation}
    p\cos\theta={\hbar\Omega}\cos\theta_{\mathrm{res}}\:/({2v}).
\end{equation}
In Eqs.~(\ref{angleinterv1})-(\ref{Lactual}) we used the smallness
of angle $\theta$.

Performing in Eq.~(\ref{Landauer}) the integration over the
intervals (\ref{momentainterv})-(\ref{angleinterv2}) of momenta and
angles and substituting the probability $P_{+1}(\bp)$ by $\cL$,
Eq.~(\ref{Lactual}), we arrive at the main results of our Brief Report,
Eqs.~(\ref{result1}) and (\ref{result2}).

{\it Physical interpretation.} Equation~(\ref{result1}) indicates that
the photocurrent quickly decreases with wavelength in the range of
photon energies exceeding the height of the potential barrier.
Indeed, only electrons with energies close to $\pm\hbar\Omega/2$
participate in the photon-assisted transport. With increasing this
energy the shallow potential becomes more transparent and has a
lesser effect on electron motion, leading to the decrease of the
photocurrent.

Since the energies of the involved electrons lie far below or far
above the Fermi level, the photocurrent weakly depends on the
background gate voltage. The dependency on the polarization
direction $\propto\sin^2\gamma$ can be understood as follows. The
photocurrent comes mainly from electrons moving nearly parallel to
the barrier, as the others' motion is unimpeded by the potential.
The light absorption rate is proportional to the square of the
component of the electric field $E_\bot$ perpendicular to the electron
velocity, where $E_\bot\propto\sin\gamma$ for the specified electrons.

Albeit the photon absorption occurs far from the Dirac point and
thus its description applies as well to an ordinary semiconductor,
the dependency of the photocurrent on the radiation frequency
$\Omega$, Eqs.~(\ref{result1}) and (\ref{result2}), is determined by
the bandstructure of graphene. Because there is no gap between the
conduction and the valence bands, the photocurrent in graphene does
not vanish even at very low frequencies, contrary to the case of an
ordinary semiconductor.

The dependency of the current on the polarization can be used to
separate the resonant photon absorption, considered here, from the
other possible mechanisms of generating the photocurrent (e.g., a
nonuniform heating of the sample by light). In the latter case
phonons may play the same role as photons, but there would be no
polarization dependency of the current.

{\it Estimation.} Let us estimate the value of the photocurrent for
the typical device parameters\cite{Xia:photoimaging}. For
$\hbar\Omega = 2 e\mathrm{V}$, $U_0 = 50 \mathrm{m}e\mathrm{V} $, $W
= 0.6 \mu \mathrm{m}$, $S = 13\mathrm{kW}/\mathrm{cm}^2$, and the
characteristic size of the potential step $L=100 \mathrm{nm}$
($F=U_0/L$) we obtain from Eq.~(\ref{result1}) at $\gamma=\pi/2$ the
current $I\approx 12 \mathrm{nA}$, in agreement with the
characteristic values of the current measured in
Ref.~\onlinecite{Xia:photoimaging}. In principle, the recombination
length of photoexcited charge carriers may be comparable with $L$,
then one should anticipate an attenuation of the current within an
order of magnitude. As we mentioned before, checking its dependency
on polarization could verify and help one to further investigate the
mechanism of the current generation.

{\it Conclusion.} We calculated photocurrent in a graphene junction
irradiated by light with the photon energy $\hbar\Omega$
considerably exceeding the characteristic height $U_0$ of the
potential. The result is significantly smaller than the photocurrent
in the case $\hbar\Omega\lesssim U_0$. It strongly depends on the
polarization of light and is weakly affected by the background gate
voltage. We thank M.V.~Fistul for discussions. Our work has been
supported financially by SFB Transregio 12 and SFB 491.


\end{document}